\documentclass[preprint,showpacs,nofootinbib,preprintnumbers,amsmath,amssymb]{revtex4}
\usepackage{mathrsfs}
\usepackage{graphicx}
\usepackage{bm}
\usepackage{slashed}

\newcommand*{\s}{\slashed}
\newcommand*{\T}{\perp}

\begin{document}

\title{Pretzelosity $h_{1T}^{\perp}$ and quark orbital angular momentum}

\author{Jun She}
\author{Jiacai Zhu}
\author{Bo-Qiang Ma}\email{mabq@phy.pku.edu.cn}
\affiliation{School of Physics and State Key Laboratory of Nuclear
Physics and Technology, Peking University, Beijing 100871, China}

\begin{abstract}
We calculate the pretzelosity distribution ($h_{1T}^{\perp}$), which
is one of the eight leading twist transverse momentum dependent
parton distributions (TMDs), in the light-cone formalism. We find
that this quantity has a simple relation with the quark orbital
angular momentum distribution, thus it may provide a new possibility
to access the quark orbital angular momentum inside the nucleon. The
pretzelosity distribution can manifest itself through the
$\sin(3\phi_h-\phi_S)$ asymmetry in semi-inclusive deep inelastic
scattering process. We calculate the $\sin(3\phi_h-\phi_S)$
asymmetry at HERMES, COMPASS, and JLab kinematics and present our
prediction on different targets including the proton, deuteron, and
neutron targets. Inclusion of transverse momentum cut in data
analysis could significantly enhance the $\sin(3\phi_h-\phi_S)$
asymmetry for future measurements.
\end{abstract}

 \maketitle

\newpage

\section{Introduction}
The nucleon is a composite system, so its spin not only comes from
the intrinsic spins of the constituents, but also from the orbital
angular momenta due to the relative motion of quarks and gluons. One
of the main tasks of hadron physics is to quantitatively know the
contribution from each component of the above items. However, on the
theoretical side, there are disputes on how to define these
quantities, and decomposing the contribution from spin and orbital
angular momentum parts in experiments is also not clear yet.

Conventionally, it seems easy to start from the QCD Lagrangian and
follow the N\"{o}ther's theorem, and to write the conserved angular
momentum as
\begin{eqnarray}
\vec{J}_{\mathrm{QCD}}&=&\int
d^3x\psi^\dag\frac{\vec{\Sigma}}{2}\psi+\int d^3x
\psi^\dag\vec{x}\times(-i\nabla)\psi\nonumber\\
&+&\int d^3x\vec{E}^a\times\vec{A}^a+\int d^3x
E^{ai}\vec{x}\times\nabla
A^{ai}\nonumber\\&\equiv&\vec{S}_q+\vec{L}_q+\vec{S}_g+\vec{L}_g,
\end{eqnarray}
in which the four terms are identified as the quark spin, quark
orbital angular momentum, gluon spin, and gluon orbital angular
momentum, respectively. This definition has been used in many
references~\cite{orbital,Brodsky01}, and it is naturally considered
as the generator of the space rotation. In Ref.~\cite{Brodsky01},
the matrix elements of this defined orbital angular momentum
operator in the light-cone representation were calculated, and the
results satisfy the conservation of the $z$ projection of angular
momentum:
\begin{eqnarray}
J^z=\sum_{i=1}^n s_i^z + \sum_{j=1}^{n-1} l_j^z.
\end{eqnarray}
But a problem is that except for the quark spin term, the other
three terms are gauge dependent, thus they have obscure physical
meanings in common situations.

By considering this, Ji suggested a gauge-invariant
expression~\cite{XDJ1997}:
\begin{eqnarray}
\vec{J}_{\mathrm{QCD}}&=&\int
d^3x\psi^\dag\frac{\vec{\Sigma}}{2}\psi+\int d^3x
\psi^\dag\vec{x}\times(-i\vec{D})\psi+\int d^3x
\vec{x}\times(\vec{E}\times\vec{B})\nonumber\\
&\equiv&\vec{S}_q+\vec{L}_q+\vec{J}_g,
\end{eqnarray}
where $\vec{D}=\nabla-ig\vec{A}$ is the covariant derivative. As
pointed in Ref.~\cite{XDJ1997}, this can be experimentally accessed
in the deep virtual Compton scattering (DVCS) process, where the
total angular momentum $\vec{J}_q\equiv \vec{S}_q+\vec{L}_q$ can be
measured, and HERMES Collaboration has reported their experimental
progresses in recent years~\cite{DVCS}. But in this definition,
unlike the quark case, there exists no gauge-invariant decomposition
of spin and orbital contributions for the gluon angular momentum
$\vec{J}_g$. However, de~T\'{e}ramond and Brodsky showed
recently~\cite{Brodsky08} that under the framework of light-front
QCD, there is a separation of the dynamics of quark and gluon
binding from the kinematics of constituent spin and internal orbital
angular momentum. More seriously, the assumed quark orbital angular
momentum operator does not obey the angular momentum algebra
$\vec{J}\times\vec{J}=i\vec{J}$, and there is an interaction term
involving also gluons. So it is a question whether such defined and
measured quantity can be viewed as the quark orbital angular
momentum.

Trying to reconcile the gauge invariance and the angular momentum
algebra, Chen {\it et al.} proposed a new form of spin
sum~\cite{XSC2008} by adding a surface term $\int
d^3x\nabla\cdot[\vec{E}^a(\vec{A}^a_{\mathrm{pure}}\times\vec{x})]$,
which vanishes after integration over the space,
\begin{eqnarray}
\vec{J}_{\mathrm{QCD}}&=&\int
d^3x\psi^\dag\frac{\vec{\Sigma}}{2}\psi+\int d^3x
\psi^\dag\vec{x}\times(-i\vec{D}_{\mathrm{pure}})\psi\nonumber\\
&+&\int d^3x\vec{E}^a\times\vec{A}_{\mathrm{phys}}^a+\int d^3x
E^{ai}\vec{x}\times\nabla
A_{\mathrm{phys}}^{ai}\nonumber\\&\equiv&\vec{S}_q+\vec{L}_q+\vec{S}_g+\vec{L}_g,
\end{eqnarray}
where $\vec{D}_{\mathrm{pure}}=\nabla-ig\vec{A}_{\mathrm{pure}}$,
and $\vec{A}_{\mathrm{pure}}$ is the pure gauge component of
$\vec{A}$, satisfying
$\vec{D}_{\mathrm{pure}}\times\vec{A}_{\mathrm{pure}}=0$. $\vec{A}$
can be decomposed as
$\vec{A}=\vec{A}_{\mathrm{pure}}+\vec{A}_{\mathrm{phys}}$. In this
definition, all terms are gauge invariant and obey the angular
momentum algebra. But there is a practical problem of lacking a
clear way to measure orbital angular momentum defined so far.

If we disregard the disputes on the definition, we can define
various quantities under each frame of definition and investigate
their properties. In Ref.~\cite{Ma1998}, the first definition
mentioned above was used and a quantity that represents the quark
orbital angular momentum in the light-cone formalism was obtained.
It was shown that the quark orbital angular momentum equals the
difference between the helicity and transversity, but it was not
clear how to measure this quantity directly at that time. In
Ref.~\cite{Avakian2008}, Avakian {\it et al.} used the bag model and
the spectator model to deduce the result that the difference of
helicity and transversity is (a transverse moment of) the so-called
pretzelosity, a new quantity which is one of the leading twist
parton distributions. Later in Ref.~\cite{Pasquini2008}, Pasquini
{\it et al.} confirmed this statement but using the light-cone
constituent quark model. Conventionally, the difference of helicity
and transversity is viewed as the relativistic effects in the
nucleon, so in Refs.~\cite{Avakian2008,Pasquini2008}, pretzelosity
is considered as the measurement of the relativistic effects in the
nucleon. In this paper, we calculate the pretzelosity distribution
in the light-cone SU(6) quark-diquark model and reconfirm the
conclusion as that in Refs.~\cite{Avakian2008,Pasquini2008}. So
combining with the result already obtained in Ref.~\cite{Ma1998}, we
suggest that pretzelosity might open a new way to access the quark
orbital angular momentum. Of course we should remember that our
definition and conclusion are based on the light-cone gauge in the
infinite momentum frame (IMF), or equivalently, $A^+=0$ gauge in the
light-cone formalism~\cite{Brodsky82,Brodsky98}, which is the
conventional language for describing parton distributions inside the
nucleon. The advantages of using light-cone formalism to describe
the quark orbital angular moment have been also stressed in
Refs.~\cite{Brodsky01,Brodsky08}.

\section{TMDs and pretzelosity}

Pretzelosity, usually denoted as $h^\T_{1T}$, is one of the eight
leading twist transverse dependent parton distributions (TMDs). Up
to the leading twist, the decomposition of the quark-quark
correlator reads~\cite{Mulders1996,Goeke2005}
\begin{eqnarray}
\Phi(x,{\bm p}_\T)=\frac{1}{2}\{f_1\s{n}_+ -
f^\T_{1T}\frac{\epsilon^{ij}_\T p_\T^i S_\T^j}{M_N}\s{n}_+
+(S_\parallel g_{1L} + \frac{{\bm p}_\T\cdot{\bm
S}_\T}{M_N}g_{1T})\gamma_5\s{n}_+ \nonumber\\
+ h_{1T}\frac{[\s{S}_\T,\s{n}_+]\gamma_5}{2} + (S_\parallel
h^\T_{1L}+\frac{{\bm p}_\T\cdot{\bm
S}_\T}{M_N}h^\T_{1T})\frac{[\s{p}_\T,\s{n}_+]\gamma_5}{2M_N} +
ih_1^\T\frac{[\s{p}_\T,\s{n}_+]}{2M_N}\}.
\end{eqnarray}
Using the abbreviation $\Phi^{[\Gamma]}=\mathrm{Tr}(\Phi\Gamma)/2$,
we can get the traces of the correlator~\cite{Mulders1996,Goeke2005}
\begin{eqnarray}
\Phi^{[\gamma^+]}&=&f_1-S_\T^j \frac{\epsilon^{ij}_\T p_\T^i}{M_N}
f^\T_{1T}
,\\
\Phi^{[\gamma^+\gamma_5]}&=&S_\parallel g_{1L} + \frac{{\bm
p}_\T\cdot{\bm
S}_\T}{M_N}g_{1T},\\
\Phi^{[i\sigma^{i+}\gamma_5]}&=&S^i_\T h_1 + S_\parallel
\frac{p^i_\T}{M_N} h^\T_{1L} + S_\T^j\frac{2p^i_\T p^j_\T-p^2_\T
\delta^{ij}_\T}{2M_N^2}h^\T_{1T} - \frac{\epsilon^{ij}_\T
p_\T^j}{M_N}h_1^\T.
\end{eqnarray}
Working in the light-cone gauge, we
have~\cite{Mulders1996,Barone2002}
\begin{eqnarray}
\Phi^{[\Gamma]}=\int \frac{d\xi^- d^2{\bm \xi}_\T}{16\pi^3}
e^{i(xP^+\xi^--{\bm p}_\T\cdot{\bm \xi}_\T)}\langle PS
|\bar{\psi}(0)\Gamma\psi(0,\xi^-,\xi_\T)|PS\rangle.\label{trace}
\end{eqnarray}
The pretzelosity distribution can be worked out by
\begin{eqnarray}
\frac{p^x_\T p^y_\T}{M_N^2}h^\T_{1T}(x,p^2_\T)=\int \frac{d\xi^-
d^2{\bm \xi}_\T}{16\pi^3} e^{i(xP^+\xi^--{\bm p}_\T\cdot{\bm
\xi}_\T)} \langle PS^y
|\bar{\psi}(0)i\sigma^{1+}\gamma_5\psi(0,\xi^-,\xi_\T)|PS^y\rangle,~~~
\label{pretzelosity}
\end{eqnarray}
where $|PS^y\rangle$ denotes the hadronic state with a polarization
in $y$ direction.

Pretzelosity is chirally odd, so it must couple with a chirally odd
partner in the semi-inclusive deep inelastic scattering (SIDIS)
process (see Sec. V). It is usually considered as a measure of the
relativistic effects in the nucleon, and a detailed discussion on
its property can be found in Ref.~\cite{Avakian2008}.

\section{Pretzelosity in the light-cone SU(6) quark-diquark model}

The matrix element shown in Eq.~(\ref{pretzelosity}) cannot be
solved strictly since a proton state contains the nonperturbative
information. Conventionally, this state can be expanded in a series
of light-cone Fock states with coefficients, i.e., the light-cone
wave functions, and all the wave functions contain the full
information of the nucleon. In principle, there is an infinite
number of wave functions and the wave functions cannot be calculated
perturbatively. However, this expansion can be truncated so that
only the Fock states with a few partons necessary for calculation
are left, and the finite number of wave functions can be
parametrized by experiments or certain models. In this paper, we use
the light-cone SU(6) quark-diquark model~\cite{Ma96}, in which the
proton state is constructed by a valence quark and a spectator
diquark. The advantage of this model is that some of the gluon and
quark effects in the spectator can be effectively described by the
diquark with a few parameters. However, as only valence quark
distributions can be directly calculated in this model, one must use
other information to take into account contributions from sea and
gluon distributions.

Any hadron state can be expanded in terms of a complete set of Fock
states at equal ``light-cone time"~\cite{Brodsky82,Brodsky98}
\begin{eqnarray}
|H\rangle=&&\sum_{n,\lambda_i}\int[dx][d^2\bm{k}_\T]\psi_n(x_i,\bm{k}_{\T
i},\lambda_i)\nonumber\\
&&\prod_{q}\frac{u(x_i P^+,x_i \bm{P}_\T+\bm{k}_{\T
i},\lambda_i)}{\sqrt{x_i}}\prod_{g}\frac{\epsilon(x_i P^+,x_i
\bm{P}_\T+\bm{k}_{\T i},\lambda_i)}{\sqrt{x_i}}|n\rangle,
\label{Fock1}
\end{eqnarray}
with the normalization condition
\begin{eqnarray}
\sum_{n,\lambda_i}\int[dx][d^2\bm{k}_\T]|\psi_n(x_i,\bm{k}_{\T
i},\lambda_i)|^2=1,\label{norm}
\end{eqnarray}
and the integral over the phase space is
\begin{eqnarray}
[dx]=\delta(1-\sum_{i=1}^n
x_i)\prod_{i=1}^{n}dx_i,~[d^2\bm{k}_\T]=16\pi^3\delta^2(\sum_{i=1}^n
\bm{k}_{\T i})\prod_{i=1}^{n}(d^2\bm{k}_{\T i}/16\pi^3).~~
\end{eqnarray}

In the SU(6) quark-diquark model, the proton state with a spin
component $S_z= \pm\frac{1}{2}$ in the instant form can be written
as
\begin{eqnarray}
|p^\uparrow\rangle&=&\frac{1}{3}\sin\theta\varphi_V\left[(ud)^0u^\uparrow
- \sqrt{2}(ud)^1u^\downarrow - \sqrt{2}(uu)^0d^\uparrow +
2(uu)^1d^\downarrow\right])+\cos\theta\varphi_S(ud)^Su^\uparrow,~~~~\label{pup}\\
|p^\downarrow\rangle&=&-\frac{1}{3}\sin\theta\varphi_V\left[(ud)^0u^\downarrow
- \sqrt{2}(ud)^{-1}u^\uparrow - \sqrt{2}(uu)^0d^\downarrow +
2(uu)^{-1}d^\uparrow\right]+\cos\theta\varphi_S(ud)^Su^\downarrow,~~~~`
\label{pdown}
\end{eqnarray}
where $\theta$ is the mixing angle that breaks the SU(6) symmetry
when $\theta\neq\pi/4$. More explicitly, we write the Fock expansion
in a diquark model as
\begin{eqnarray}
|PS^\uparrow\rangle=&&\sum_j\frac{1}{16\pi^3}\int\frac{dx_qdx_D}{\sqrt{x_qx_D}}\int
d^2\bm{k}_{q\T}d^2\bm{k}_{D\T},\nonumber\\
&&\delta(1-x_q-x_D)\delta^2(\bm{k}_{q\T}+\bm{k}_{D\T})\psi(x_j,\bm{k}_{j\T},\lambda_i)a_j^\dag
b_j^\dag|0\rangle,~~~ \label{Fock2}
\end{eqnarray}
where the wave functions can be extracted from Eq.~(\ref{pup}). The
negative helicity state can be written in the same way.

Notice that now we are working in the instant frame, and we need to
transform to the light-cone frame. The connection between the two
frames is through a Melosh-Wigner rotation~\cite{Melosh74,Ma91}. For
a spin-$\frac{1}{2}$ particle, we use $q_T$ and $q_F$ to denote the
instant and light-cone spinors, respectively, and the Melosh-Wigner
rotation is known to be~\cite{Ma91}
\begin{eqnarray}
\left(\begin{array}{c}
q_F^\uparrow\\
q_F^\downarrow
\end{array}\right)
=\omega\left(\begin{array}{cc}
k^++m & -k^R\\
k^L & k^++m
\end{array}\right)\left(\begin{array}{c}
q_T^\uparrow\\
q_T^\downarrow
\end{array}\right)
\equiv \bm{M}^{1/2}\left(\begin{array}{c}
q_T^\uparrow\\
q_T^\downarrow
\end{array}\right), \label{Melosh}
\end{eqnarray}
where $k^{R,L}=k_1\pm ik_2$,
$\omega=\left[(x\mathscr{M}_D+m_q)^2+p^2_\T\right]^{-1/2}$ with
$\mathscr{M}_D^2=\frac{m_q^2+p^2_\T}{x}+\frac{m_D^2+p^2_\T}{1-x}$,
$m_q$ and $m_D$ are mass parameters for the quark and diquark, and
$\bm{M}^{1/2}$ denotes the Melosh-Wigner rotation matrix with the
superscript referring to the spin of the particle. For the spin-0
scalar diquark, there is no such transformation. For the spin-1
vector diquark, the transformation is represented by a $3\times3$
matrix $\bm{M}^1$, whose explicit expression can also be found in
Ref.~\cite{Ma2002}. In practice, we will not use the explicit form,
for it does not appear in the final expression for the spectator
debris due to the unitary property $\bm{M}^{1\dag}\bm{M}^1=1$.

Using Eqs.~(\ref{trace}), (\ref{Fock1}), (\ref{Fock2}), and
(\ref{Melosh}), we deduce the formula for calculating the trace of
the correlator:
\begin{eqnarray}
\phi^{[\Gamma]}=&&\sum_{j,\lambda,\lambda^\prime,\lambda_D}
\frac{1}{32\pi^3}\frac{1}{xP^+}\psi_j^\ast(x,\bm{p}_\T,\lambda;1-x,-\bm{p}_\T,\lambda_D)
\psi_j(x,\bm{p}_\T,\lambda^\prime;1-x,-\bm{p}_\T,\lambda_D)\nonumber\\
&&\bar{u}(xP^+,\bm{p}_\T,\lambda)\Gamma
u(xP^+,\bm{p}_\T,\lambda^\prime).
\end{eqnarray}
We substitute $i\sigma^{1+}\gamma_5$ for $\Gamma$ and notice that
$|P^y\rangle=\frac{1}{\sqrt{2}}(|P^\uparrow\rangle+i|P^\downarrow\rangle)$,
after a careful calculation, we get the result for pretzelosity (the
superscript $v$ denotes that it is valid only for valence quarks):
\begin{eqnarray}
\label{pre}
h^{\T(uv)}_{1T}(x,\bm{p}_\T)&=&-\frac{1}{16\pi^3}\times(\frac{1}{9}\sin^2\theta\varphi_V^2W_V
-\cos^2\theta\varphi_S^2W_S),\nonumber\\
h^{\T(dv)}_{1T}(x,\bm{p}_\T)&=&-\frac{1}{8\pi^3}\times\frac{1}{9}\sin^2\theta\varphi_V^2W_V,
\end{eqnarray}
with the Melosh-Wigner rotation factor $W_D(D=V,S)$ given by
\begin{eqnarray}
W_D(x,\bm{p}_\T)=-\frac{2M_N^2}{(x\mathscr{M}_D+m_q)^2+p^2_\T},
\end{eqnarray}
which agrees with the result in Ref.~\cite{Pasquini2008}, in which a
three quark model was used. $\varphi_V$($\varphi_S$) are the wave
functions in the momentum space for the vector(scalar) diquark,
which can be parametrized by the Brodsky-Huang-Lepage (BHL)
prescription~\cite{Brodsky82,Huang1994}:
\begin{eqnarray}
\varphi_D(x,\bm{p}_\T)=A_D\exp\{-\frac{1}{8\alpha_D^2}
[\frac{m_q^2+{p}_\T^2}{x}+\frac{m_D^2+{p}_\T^2}{1-x}]\}, \label{bhl}
\end{eqnarray}
whose normalization should be consistent with Eq.~(\ref{norm}). The
parameters such as $\alpha_D$, the quark mass $m_q$ and the diquark
mass $m_D$ can be found in Ref.~\cite{Ma96}, and we list them in
Table~\ref{parameter}.
\begin{table}
\caption{\label{parameter}Parameters}
\begin{tabular}{cccc}
\hline \hline
$\alpha_D$  & $m_q$  & $m_S$  &$m_V$  \\
(MeV)&(MeV)&(MeV)&(MeV)\\
\hline
330&330&600&800\\
\hline\hline
\end{tabular}
\end{table}
In the same way, the unpolarized distributions can be
derived~\cite{Ma96}
\begin{eqnarray}
\label{unp}
f_1^{(uv)}(x,\bm{p}_\T)&=&\frac{1}{16\pi^3}\times(\frac{1}{3}\sin^2\theta\varphi_V^2
+\cos^2\theta\varphi_S^2),\nonumber\\
f_1^{(dv)}(x,\bm{p}_\T)&=&\frac{1}{8\pi^3}\times\frac{1}{3}\sin^2\theta\varphi_V^2.
\end{eqnarray}
Using above formulas, we can calculate pretzelosity and the
unpolarized distribution function independently, but this model
calculation strongly depends on the choice of the wave function. The
BHL form we use exhibits the sharp falloff at both the large and
small $x$ region, but the consequent result for the distribution
functions might be inconsistent with the realistic situation. For
example, the CTEQ6 extraction shows a divergence tendency at $x
\rightarrow 0$, rather than $0$ as Eq.~(\ref{unp}) indicates.

In order to make our result more close to the realistic situation,
we could adopt another prescription based on the results of our
model. This can be done by combining Eqs.~(\ref{pre}) and
Eq.~(\ref{unp}), then we get the relation between the unpolarized
and pretzelosity distributions:
\begin{eqnarray}
\label{pre_para}
h^{\T(uv)}_{1T}(x,\bm{p}_\T)&=&\left[f_1^{(uv)}(x,\bm{p}_\T)
-\frac{1}{2}f_1^{(dv)}(x,\bm{p}_\T)\right]W_S(x,\bm{p}_\T)
-\frac{1}{6}f_1^{(dv)}(x,\bm{p}_\T)W_V(x,\bm{p}_\T),\nonumber\\
h^{\T(dv)}_{1T}(x,\bm{p}_\T)&=&-\frac{1}{3}f_1^{(dv)}(x,\bm{p}_\T)W_V(x,\bm{p}_\T).
\end{eqnarray}
Now, we will use the phenomenological extraction of the unpolarized
distribution functions as an input to calculate the pretzelosity
distribution. For example, we will use the CTEQ6L~\cite{cteq}
parametrization to get $f_1(x)$, which have been well tested and
constrained by many experiments. And then by combining with some
assumed transverse momentum $\bm{p}_\T$ factor one can get
phenomenological $f_1(x,\bm{p}_\T)$ as input to calculate the
pretzelosity distribution $h^{\T}_{1T}(x,\bm{p}_\T)$.

Here we need to make some comments on Eq.~(\ref{pre_para}). The
left-hand side of the equation is a chiral-odd function, which does
not have gluon counterparts. But the right-hand side is a linear
combination of two chiral-even functions, which have gluon
counterparts. So the two sides have different evolutions, which
means that the relation cannot be held exactly at any scale. This
paradox originates from the fact that the model we use is a no-gluon
model, which was pointed out in Ref.~\cite{Avakian2008}.
Nevertheless, we can comprehend this relation from two aspects.
First, we assume that Eq.~(\ref{pre_para}) is valid at an initial
scale. Second, the evolution effect for pretzelosity appearing on
the left is partially (not all) contained in the unpolarized
distribution. So we conclude that Eq.~(\ref{pre_para}) could be
approximately satisfied and useful for estimating pretzelosity, but
we must be careful that the valid scale should not be too large.

The two approaches to calculate the pretzelosity distribution may
lead to different results. We denote the former approach, i.e., the
model calculation with wave function as input, as approach 1, and
the latter one, i.e., with phenomenological parametrization of
$f_1(x,\bm{p}_\T)$ as input, as approach 2. Here we should emphasize
again that we have used a ``valence model,'' and we can only
directly calculate valence quark distributions in principle. In
approach 1, for both pretzelosity and unpolarized distribution, only
the contribution from valence quarks can be considered due to a
model calculation. In approach 2, the prescription for pretzelosity
is the same as that in approach 1, but for unpolarized distribution,
the phenomenological parametrization can provide us with the
information on sea quarks so that we can take into account the sea
contributions for the unpolarized processes.  In this paper, we will
use both approaches to obtain distribution functions and give
predictions separately. Numerical results will be presented in Sec.
V.

\section{Pretzelosity and orbital angular momentum}

The effect of Melosh-Wigner rotation is also important in other
leading twist distribution functions such as the helicity and
transversity distributions, which have been extensively discussed in
light-cone formalism in Refs.~\cite{Ma96,Ma_PLB1998}. We find that
all the distributions have similar form as Eq.~(\ref{pre_para})
shows, but with different Melosh-Wigner rotation factors. The
rotation factors for the helicity and transversity distributions are
$\frac{(x\mathscr{M}_D+m_q)^2-p^2_\T}{(x\mathscr{M}_D+m_q)^2+p^2_\T}$
\cite{Ma96,Ma91} and
$\frac{(x\mathscr{M}_D+m_q)^2}{(x\mathscr{M}_D+m_q)^2+p^2_\T}$
\cite{Ma_PLB1998}, respectively. Note the simple relation
\begin{eqnarray}
\frac{(x\mathscr{M}_D+m_q)^2-p^2_\T}{(x\mathscr{M}_D+m_q)^2+p^2_\T}
-\frac{(x\mathscr{M}_D+m_q)^2}{(x\mathscr{M}_D+m_q)^2+p^2_\T}
=\frac{{p}^2_\T}{2M_N^2}W_D(x,\bm{p}_\T),
\end{eqnarray}
so we can immediately yield the equation
\begin{eqnarray}
h^{\T(1) qv}_{1T}(x, \bm{p}_\T)\equiv\frac{{p}^2_\T}{2M_N^2}h^{\T
qv}_{1T}(x, \bm{p}_\T)=g_1^{qv}(x, \bm{p}_\T)-h_1^{qv}(x,
\bm{p}_\T),
\end{eqnarray}
where $g_1$ and $h_1$ denote the helicity and transversity
distributions, respectively. This relation has already been obtained
in Ref.~\cite{Avakian2008} with a bag model and a spectator model,
and also in Ref.~\cite{Pasquini2008} with a three-quark model. But
in Ref.~\cite{Bacchetta2008}, this relation is not fully satisfied
in a spectator model where the axial-vector coupling will spoil the
relation. We think that this issue needs further discussing.

Now, we should point out that the Melosh-Wigner rotation factor for
$h^{\T(1) qv}_{1T}(x, \bm{p}^2_\T)$ is
$\frac{-p^2_\T}{(x\mathscr{M}_D+m_q)^2+p^2_\T}$, which is nothing
else but the rotation factor $M_L(x,\bm{p}_\T)$ with a minus sign
introduced in Ref.~\cite{Ma1998}. More importantly, it is the
rotation factor for the quark orbital angular momentum indicated in
Ref.~\cite{Ma1998}, so we see that there is a simple relation
between the pretzelosity and the quark orbital angular momentum
\begin{eqnarray}
L^{qv}(x,\bm{p}_\T)=-h^{\T(1) qv}_{1T}(x, \bm{p}_\T) =h_1^{qv}(x,
\bm{p}_\T)-g_1^{qv}(x, \bm{p}_\T),\label{L1}
\end{eqnarray}
or at the integration level
\begin{eqnarray}
L^{qv}(x)=\int d^2\bm{p}_\T L^{qv}(x,\bm{p}_\T)=-h^{\T(1)
qv}_{1T}(x)=h_1^{qv}(x)-g_1^{qv}(x). \label{L2}
\end{eqnarray}
We should mention that the orbital angular momentum we denote here
is under the definition that the quark orbital angular momentum
operator is $\vec{x}\times(-i\nabla)$, and the gauge we choose is
$A^+=0$.

\section{Predictions on the $\sin(3\phi_h-\phi_S)$ asymmetry
at HERMES, COMPASS and JLab kinematics}

As we pointed out before the pretzelosity has a simple relation with
the quark orbital angular momentum, thus a measurement of
pretzelosity may reveal the information on the quark orbital angular
momentum. Fortunately, the pretzelosity distribution can be measured
through $\sin(3\phi_h-\phi_S)$ asymmetry in the SIDIS
process~\cite{Kotzinian1995,Bacchetta2007}, where the cross section
can be written as
\begin{eqnarray}
\frac{d^6\sigma_{UT}}{dxdyd\phi_Sdzd^2\bm{P}_{h\T}}=&&\frac{2\alpha^2}{sxy^2}\{
(1-y+\frac{1}{2}y^2)F_{UU}
+S_\T\sin(3\phi_h-\phi_S)(1-y)F^{\sin(3\phi_h-\phi_S)}_{UT}\nonumber\\
&&+\text{other structure functions}\},
\end{eqnarray}
with
\begin{eqnarray}
&&F_{UU}=\mathcal{F}[f_1 D_1],\\
&&F^{\sin(3\phi_h-\phi_S)}_{UT}=\mathcal{F}[\frac{2(\hat{\bm{h}}\cdot\bm{p}_\T)
(\bm{p}_\T\cdot\bm{k}_\T) +\bm{p}_\T^2(\hat{\bm{h}}\cdot\bm{k}_\T)-
4(\hat{\bm{h}}\cdot\bm{p}_\T)^2(\hat{\bm{h}}\cdot\bm{k}_\T)}{2M_N^2M_h}
h_{1T}^\T H_1^\T],~~~~~
\end{eqnarray}
where a compact notation
\begin{eqnarray}
\mathcal{F}[\omega f D]=\sum_q e_q^2\int d^2\bm{p}_\T d^2\bm{k}_\T
\delta^2(\bm{p}_\T-\bm{k}_\T-\bm{P}_{h\T}/z)\nonumber\\
\omega(\bm{p}_\T,\bm{k}_\T)f^q(x,\bm{p}_\T^2)D^q(z,z^2\bm{k}_\T^2)
\end{eqnarray}
is used. Then we obtain the $\sin(3\phi_h-\phi_S)$ asymmetry
\begin{eqnarray}
A_{UT}^{\sin(3\phi_h-\phi_S)}=\frac{\frac{2\alpha^2}{sxy^2}
(1-y)F^{\sin(3\phi_h-\phi_S)}_{UT}}{\frac{2\alpha^2}{sxy^2}(1-y+\frac{1}{2}y^2)F_{UU}}.
\end{eqnarray}

First, we will present the results for the ratio of (the first
moment of) pretzelosity (i.e. $L^q(x)$ equivalently) and unpolarized
distributions, in both approaches. As we mentioned above, we can
only calculate valence quarks distributions in approach 1, whereas
in approach 2 we can take into account the sea quark contribution in
$f_1(x)$. For approach 1, we can use Eqs.~(\ref{pre})-(\ref{bhl})
directly to calculate. For approach 2, we adopt the CTEQ6L
parametrization~\cite{cteq} for the unpolarized distribution in this
paper and assume a Gaussian form factor of transverse momentum as
suggested in Ref.~\cite{Anselmino2005}:
\begin{eqnarray}
f_1(x,\bm{p}_\T)=f_1(x)\frac{\exp(-{p}_\T^2/p_{av}^2)}{\pi
p_{av}^2},
\end{eqnarray}
with $p_{av}^2=0.25~\mathrm{GeV}^2$.

\begin{figure}
\center
\includegraphics[scale=0.8]{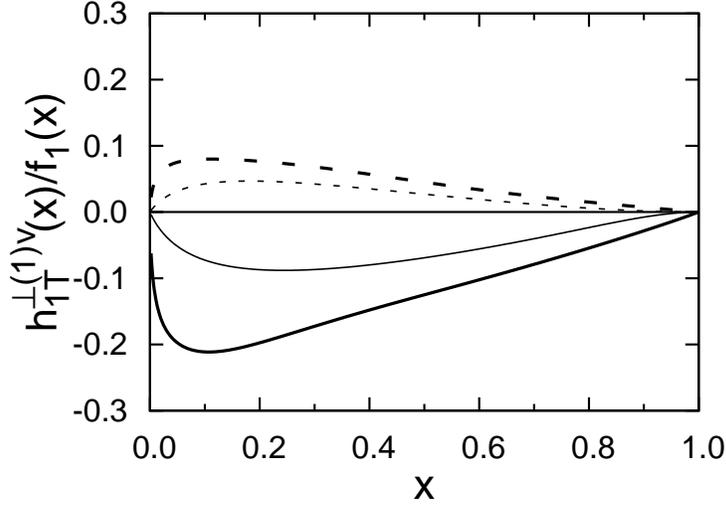}
\caption{The ratio $h^{\T(1) qv}_{1T}(x)/ f_1^q(x)$. Solid curves
for the $u$ quark and dashed curves for the $d$ quark. Thin curves
correspond to approach 1, while thick curves correspond to approach
2.} \label{plot_comp}
\end{figure}

Results are shown in Fig.~\ref{plot_comp}, with a fixed angle
$\theta=\pi/4$ and an input scale at $Q^2=4~\mathrm{GeV}^2$. At
first sight it seems strange that the ratio of pretzelosity versus
unpolarized distribution is small in approach 1 compared to that in
approach 2, as there is no sea contribution in the denominator of
approach 1. We can make a brief analysis by taking $d$ quark as an
example and we could ignore sea quarks for simplicity first. In
approach 1, we have
\begin{eqnarray}
\textrm{I}=\frac{h^{\T(dv)}_{1T}(x)}{f_1^{dv}(x)}&=&\frac{1}{3}\frac{\int
d^2 \bm{p}_\T \varphi_V^2(x,p^2_\T) W_V(x,p^2_\T)}{\int d^2
\bm{p}_\T
\varphi_V^2(x,p^2_\T)}\nonumber\\
&=&\frac{1}{3}\frac{\int d^2 \bm{p}_\T \exp\{-\frac{1}{4\alpha_D^2}
[\frac{m_q^2+{p}_\T^2}{x}+\frac{m_D^2+{p}_\T^2}{1-x}]\}
W_V(x,p^2_\T)}{\int d^2 \bm{p}_\T \exp\{-\frac{1}{4\alpha_D^2}
[\frac{m_q^2+{p}_\T^2}{x}+\frac{m_D^2+{p}_\T^2}{1-x}]\}}\nonumber\\
&=&\frac{1}{3}\int
dt~e^{-t}W_V(x,4\alpha_D^2x(1-x)t),~~(t=\frac{{p}_\T^2}{4\alpha_D^2x(1-x)}).
\end{eqnarray}
In approach 2, we have
\begin{eqnarray}
\textrm{II}=\frac{h^{\T(dv)}_{1T}(x)}{f_1^{dv}(x)}&=&\frac{1}{3}\frac{\int
d^2 \bm{p}_\T d_v^\textrm{cteq}(x,p^2_\T) W_V(x,p^2_\T)}{\int d^2
\bm{p}_\T
d_v^\textrm{cteq}(x,p^2_\T)}\nonumber\\
&=&\frac{1}{3}\int d^2 \bm{p}_\T \frac{\exp(-{p}_\T^2/p_{av}^2)}{\pi
p_{av}^2} W_V(x,p^2_\T)\nonumber\\
&=&\frac{1}{3}\int
dt~e^{-t}W_V(x,p_{av}^2t),~~(t=\frac{{p}_\T^2}{p_{av}^2}).
\end{eqnarray}
We can easily prove that $W_D(x,p^2_\T)$ is an increasing function
of $p^2_\T$ and using the inequality
\begin{eqnarray}
4\alpha_D^2x(1-x)\leq\alpha_D^2< p_{av}^2,
~~(\alpha_D=330\mathrm{MeV},~p_{av}=500\mathrm{MeV})
\end{eqnarray}
we get that I$<$II, which means that two ratios can be quite
different due to different parametrizations. Even if we consider the
sea quarks in the denominator, II will be suppressed, but still can
be larger than I. A similar analysis can be applied to the $u$ quark
though it is a little more complicated, and we can argue that it is
possible that II is larger than I. From the above analysis we find
the different transverse momentum dependence of quark distribution
functions may cause a big difference in model predictions. This
suggests the necessity to include transverse momentum dependence in
data analysis. Despite the differences, both approaches give the
prediction that the ratio $h^{\T(1) qv}_{1T}(x)/ f_1^{q}(x)$ is
small, which indicates that the relativistic effect, or the quark
orbital angular momentum as we suggested in our paper, is not
significant, if we will integrate the transverse momentum in data
analysis.

Before calculating the asymmetry, we still have to know the form for
fragmentation functions. For the ordinary fragmentation function, we
also adopt the Gaussian ansatz suggested in
Ref.~\cite{Anselmino2005}
\begin{eqnarray}
D_1(z,z^2\bm{k}^2_\T)=D_1(z)\frac{\exp(-z^2{k}_\T^2/R^2)}{\pi R^2},
\end{eqnarray}
with $R^2=0.2~\mathrm{GeV}^2$, and the parametrization for $D_1(z)$
can be found in Ref.~\cite{Kretzer}. The Collins function we use in
this paper was also given by Anselmino {\it et
al.}~\cite{Anselmino2008}, which is also a Gaussian function in
fact.

Next, we will present the numerical results of
$\sin(3\phi_h-\phi_S)$ asymmetry in the SIDIS at different
kinematics. For HERMES experiments, only the result on the proton
target is calculated, while for COMPASS experiments, the proton,
deuteron, and neutron targets are all assumed. As to Jefferson Lab
(JLab) experiments, the result on the proton and neutron\footnote{In
fact, a $^3$He target is used to extract neutron data in JLab
experiments.} targets are presented. The kinematics are shown in
Table~\ref{kin}.

\begin{table}
\caption{\label{kin}Kinematics at HERMES, COMPASS and JLab.}
\begin{tabular}{c|c|c}
\hline
HERMES                    & COMPASS                  & JLab                       \\
\hline \hline
$p_\mathrm{lab}=27.6~\text{GeV}$ & $p_\mathrm{lab}=160~\text{GeV}$ & $p_\mathrm{lab}=12~\text{GeV}$  \\
\hline
$Q^2>1~\text{GeV}^2$      & $Q^2>1~\text{GeV}^2$     & $Q^2>1~\text{GeV}^2$       \\
\hline
$W^2>10~\text{GeV}^2$     & $W^2>25~\text{GeV}^2$    & $W^2>4~\text{GeV}^2$       \\
\hline
$0.023<x<0.4$             &                          & $0.1<x<0.6$                \\
\hline
$0.1<y<0.85$              & $0.1<y<0.9$              & $0.4<y<0.85$               \\
\hline
$0.2<z<0.7$               & $0.2<z<1$                & $0.4<z<0.7$                \\
\hline
\end{tabular}
\end{table}

\begin{figure}
\center
\includegraphics[scale=0.7]{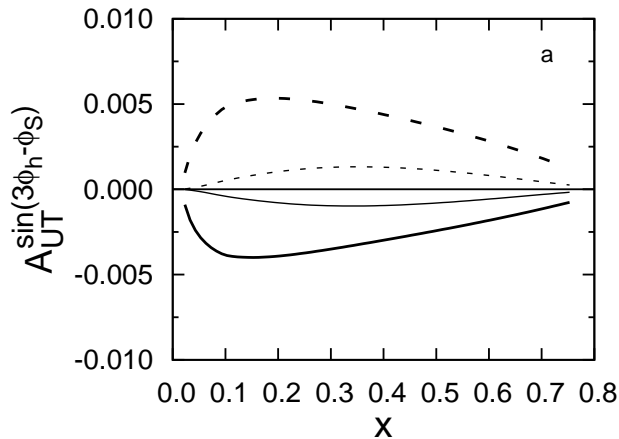}
\caption{$\sin(3\phi_h-\phi_S)$ asymmetry as a function of $x$ at
HERMES kinematics. Solid curves for the $\pi^+$ production and
dashed curves for the $\pi^-$ production. Thin curves correspond to
approach 1, while thick curves correspond to approach 2. Only the
proton target is assumed here.} \label{HERMES}
\end{figure}

\begin{figure}
\center
\includegraphics[scale=0.7]{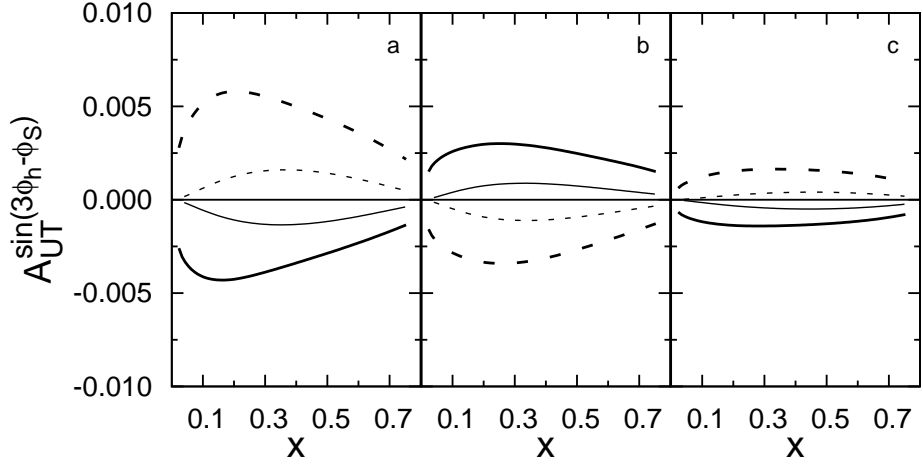}
\caption{Same as Fig.~\ref{HERMES}, but at COMPASS kinematics. A for
proton target, B for neutron target, and C for deuteron target.}
\label{COMPASS}
\end{figure}

\begin{figure}
\center
\includegraphics[scale=0.7]{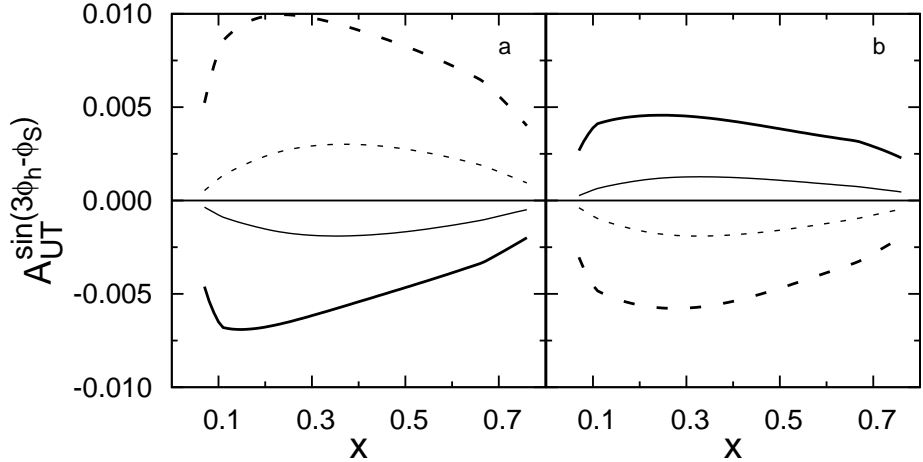}
\caption{Same as Fig.~\ref{HERMES}, but at JLab kinematics. A for
proton target and B for neutron target.} \label{JLab}
\end{figure}

The results are shown in Figs.~\ref{HERMES} - Fig.~\ref{JLab}.
Obviously, different approaches to distribution functions give quite
different predictions, and this can be understood from
Fig.~\ref{plot_comp}. Similarly, when we calculate the asymmetry,
only valence quarks are summed over in approach 1, whereas in
approach 2, all the flavors are summed over including the sea quarks
in the denominator. Perhaps approach 2 might be closer to the
realistic situation, because the parametrization has been well
constrained by many experiments and can fit the current data much
better than a simple model calculation.

Besides the difference, the two approaches also have something in
common. First, we find that both approaches predict that the
asymmetries are of different sign for $\pi^+$ and $\pi^-$
productions. Second, we notice that the asymmetry decreases as $x$
increases in the valence region, which is quite different from some
other known effects, e.g., the $\sin(\phi_h+\phi_S)$ and
$\sin(\phi_h-\phi_S)$ asymmetries. This can be explained by the
ratio of the pretzelosity and unpolarized distribution shown in
Fig.~\ref{plot_comp}. Another important feature that might be
concerned by experiments is that the asymmetry is small, up to a
maximum less than 1\%. Here taking approach 2 as an example, we could make
a comparison of our result with transversity and corresponding
$\sin(\phi_h+\phi_S)$ asymmetry, which have been intensively studied
in recent years. These two asymmetries can be simply written as
\begin{eqnarray}
A^{\sin(3\phi_h-\phi_S)}_{UT}\sim\frac{\omega_1 h_{1T}^{\T(1)}
H^{\T(1/2)}_1}{f_1 D_1},~~~
A^{\sin(\phi_h+\phi_S)}_{UT}\sim\frac{\omega_2 h_1
H^{\T(1/2)}_1}{f_1 D_1}.
\end{eqnarray}
The ratio for $h^{\T(1)}_{1T}/f_1$ has been shown in
Fig.~\ref{plot_comp}, and that for $h_1/f_1$ can be found in
Ref.~\cite{Ma_PRD2002}. We can clearly find that compared with
transversity, (the first moment of) pretzelosity is suppressed by
$1/2-1/10$. As to the oscillation function
$\omega(\bm{p}_\T,\bm{k}_\T)$, we could also make an estimation on
its effect by integrating the angle dependence of the two vectors
$\bm{p}_\T$ and $\bm{k}_\T$ first,
\begin{eqnarray}
\frac{\tilde{\omega}_1}{\tilde{\omega}_2}\sim\frac{\int d\phi_p
d\phi_k|P_{h\T}|^3\omega_1(\bm{p}_\T,\bm{k}_\T)}{\int d\phi_p
d\phi_k|P_{h\T}|^3\omega_2(\bm{p}_\T,\bm{k}_\T)}
=\frac{3p_\T^2}{2p_\T^2+k_\T^2}, \label{ratio}
\end{eqnarray}
where we multiply $\omega$ by a weight $|P_{h\T}|^3$ just for
simplifying the estimation. For a rough estimation, we assume
$\langle p_\T^2 \rangle\approx p_{av}^2=0.25\mathrm{GeV}^2, ~\langle
k_\T^2 \rangle\approx R^2/\langle z \rangle^2=0.25/\langle z
\rangle^2\mathrm{GeV}^2$, where $\langle z \rangle=0.36$ in HERMES.
We find that the ratio is about $0.3\sim0.4$. So from the above
analysis, the $\sin(3\phi_h-\phi_S)$ asymmetry is suppressed by an
order or more compared with the $\sin(\phi_h+\phi_S)$ asymmetry.
According to the HERMES data~\cite{HERMES}, the size of
$\sin(\phi_h+\phi_S)$ asymmetry is about a few percent, so our
result for a small $\sin(3\phi_h-\phi_S)$ asymmetry is reasonable
from the above expectation. Such small asymmetry seems to be
consistent with the preliminary COMPASS data~\cite{COMPASS}, thus it
would be a challenge for further experiments to measure pretzelosity
without transverse momentum information.

\begin{figure}
\center
\includegraphics[scale=0.7]{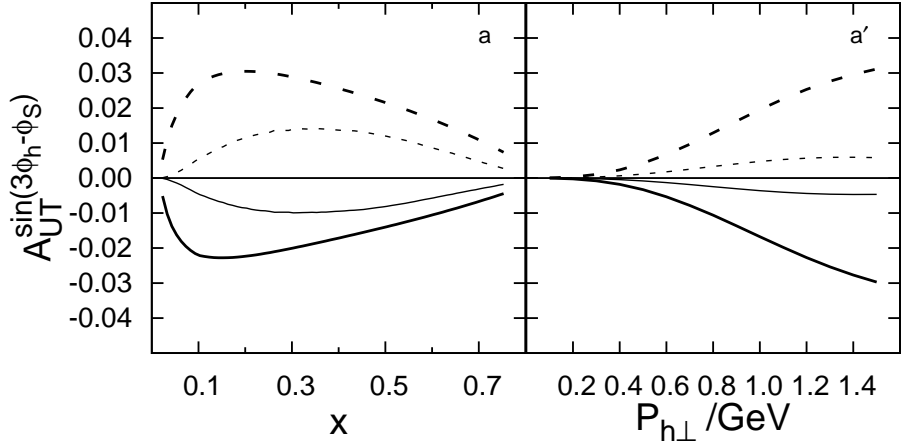}
\caption{The left panel shows the $x$ dependence of the asymmetry
with a cutoff $P_{h\T}>1.0~\mathrm{GeV}$, while the right panel
shows the $P_{h\T}$ dependence of the asymmetry after integrating
all the other kinematic variables. The curves used here are the same
as that in Fig.~\ref{HERMES}} \label{HERMES_cut}
\end{figure}

\begin{figure}
\center
\includegraphics[scale=0.7]{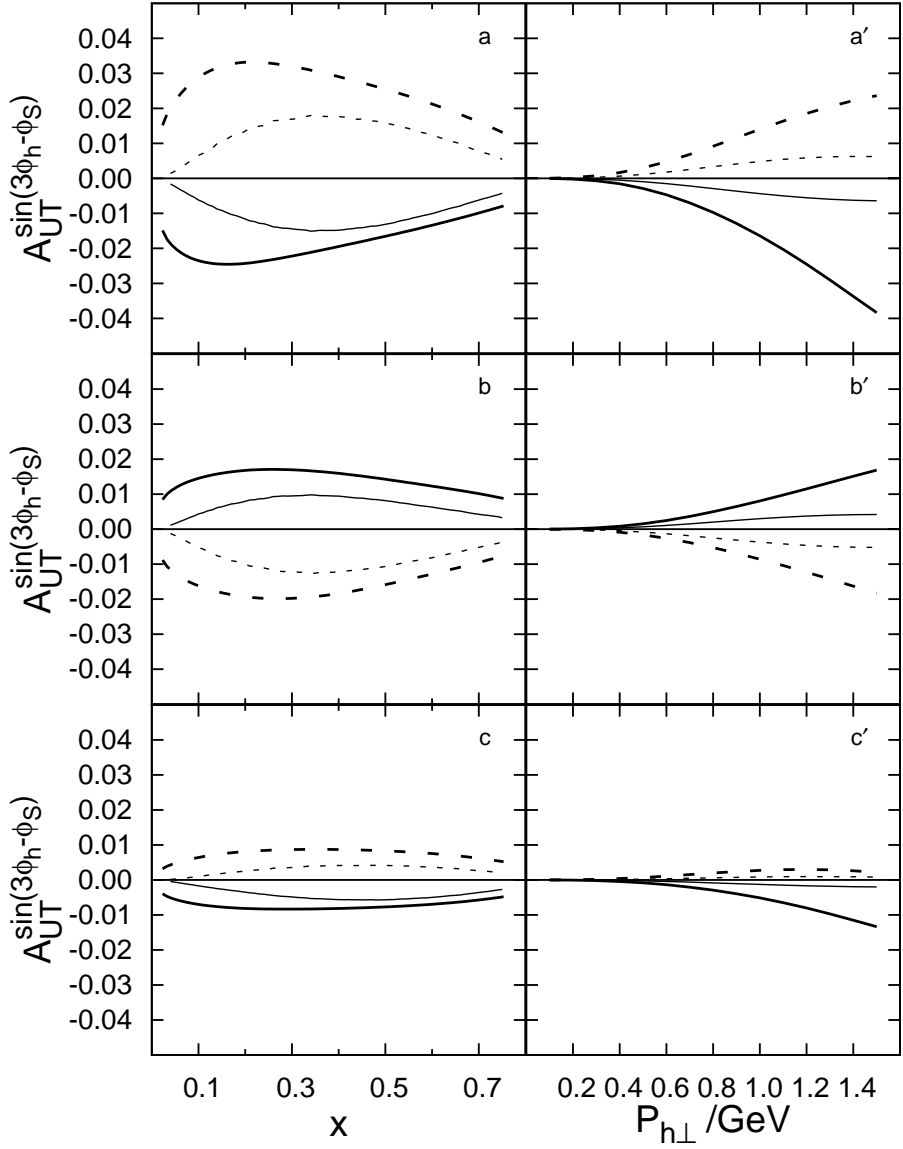}
\caption{Same as Fig.~\ref{HERMES_cut}, but at COMPASS kinematics.
The upper, middle, and lower panels correspond to the proton,
neutron, and deuteron target, respectively.} \label{COMPASS_cut}
\end{figure}

\begin{figure}
\center
\includegraphics[scale=0.7]{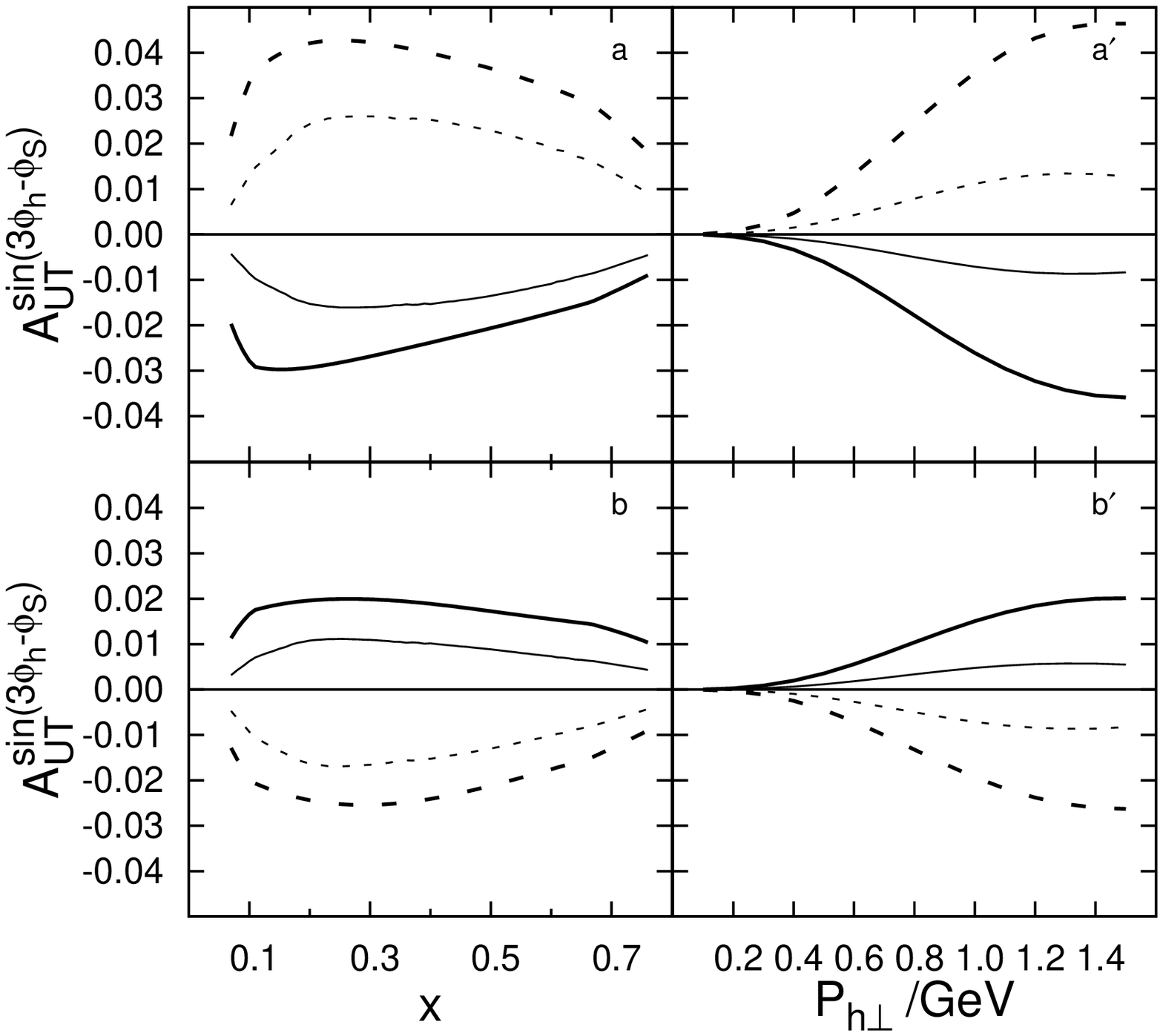}
\caption{Same as Fig.~\ref{HERMES_cut}, but at JLab kinematics. The
upper and lower panels correspond to the proton and neutron target,
respectively.} \label{JLab_cut}
\end{figure}

However, the inclusion of transverse momentum dependence in data
analysis may provide a viable way to extract pretzelosity from
$\sin(3\phi_h-\phi_S)$ asymmetry measurements. We observe that both
the Melosh-Wigner rotation factor for $h_{1T}^{\T (1)}$ and the
ratio as Eq.~(\ref{ratio}) shows are increasing functions of
$p_\T^2$, which enlightens us that selecting large $p_\T$ events in
data analysis might enhance the asymmetry. Unfortunately, the
intrinsic parton momentum $p_\T$ cannot be directly manipulated
during the measurements. A compromise method is to select large
$P_{h\T}$ events instead, for $P_{h\T}$ can be directly measured. In
this prescription, we can exclude most small $p_\T$ events and
ensure that most large $p_\T$ events will come into data analysis.
Next we will recalculate our results with a cutoff
$P_{h\T}>1.0\mathrm{GeV}$, meanwhile, we will investigate the
$P_{h\T}$ dependence of the asymmetry. The results are shown in
Fig.\ref{HERMES_cut} - Fig.\ref{JLab_cut}.
As we expect, the $x$ dependence of the asymmetry is indeed enhanced
by a few times, thus it might be measurable for the experiments,
though high statistics would be required due to the cutoff. The
$P_{h\T}$ dependence of the asymmetry is also presented, and it can
reach as large as several percent, so it provides an auxiliary
measurable quantity to extract pretzelosity. Although our proposal
to consider to cut off the small $P_{h\T}$ events will enhance the
asymmetry, we should be aware about this disposal. Our calculation
is a leading order approximation and based on the factorization for
the SIDIS, but the factorization was proved to be valid only in the
region $\Lambda_\textrm{QCD}\ll P_{h\T}\ll Q$~\cite{factorization}.
The ideal kinematic regime to study transverse momentum dependent
parton distributions is $P_{h\T}\sim\Lambda_\textrm{QCD}$ and not
too large $Q^2$. Otherwise, the gluon radiation will be important,
and a higher order pQCD correction will contribute. As pointed out
in Ref.~\cite{Anselmino2007}, this transition point is around
$P_{h\T}\approx 1\textrm{GeV}$, and the parameters such as the
Gaussian widths we used are suitable at
$P_{h\T}\leqslant1\textrm{GeV}$. So we must be careful about our
extension analysis to a larger $P_{h\T}$, and here we assume that
our results at a little larger $P_{h\T}$ but not too larger than
1GeV are still acceptable. For the experiments, they can choose an
appropriate cutoff for convenience, and meanwhile we suggest an
upper limit smaller than 2GeV for $P_{h\T}$. Another problem is that
the events will be strongly suppressed at $P_{h\T}\gg
\Lambda_\textrm{QCD}$, so it is a challenge for the experiments to
collect more data. Anyway, we expect further experiments will bring
us more information about it.

\section{Summary}

How to separate the nucleon spin into the intrinsic spin and orbital
angular momentum parts of the constituents is a fundamental but
difficult task. We have shown some debates on defining the quark
orbital angular momentum in the beginning of our paper and pointed
out that each definition has its advantage and disadvantage.
However, disregarding the rationality of the theoretical definition,
we can obtain the quark orbital angular momentum under each
definition, although they might not be the real desired quantity. In
Ref.~\cite{Ma1998}, where the expression
$\psi^\dag\vec{x}\times(-i\nabla)\psi$ was used as the definition, a
quantity $L(x)$ representing the quark orbital angular momentum in
the light-cone gauge was obtained. In this paper, also working in
the light-cone representation, we found that $L(x)$ has a simple
relation with the so-called pretzelosity distribution.

Pretzelosity is one of the eight leading twist TMD functions, whose
feature has been discussed in Ref.~\cite{Avakian2008}. It is known
that pretzelosity is considered as the difference between the
helicity and transversity distributions, reflecting the relativistic
effect of the spin structure. But according to Ref.~\cite{Ma1998},
the moment of pretzelosity is nothing but the quark orbital angular
momentum. So it provides a new possibility to access the quark
orbital angular momentum through pretzelosity, and fortunately,
pretzelosity can be measured in SIDIS through $\sin(3\phi_h-\phi_S)$
asymmetry. In our paper, under the framework of the SU(6)
quark-diquark model, we use two approaches to calculate
pretzelosity, and we present our numerical prediction on the
$\sin(3\phi_h-\phi_S)$ asymmetry under three kinematics for HERMES,
COMPASS, and JLab experiments. If we do not apply any extra
constrained condition, our prediction shows a small asymmetry that
might bring a great challenge for experiments. However, after
applying a cutoff on $P_{h\T}$, we get an enhanced asymmetry which
is measurable for experiments, though we should still be careful
that $P_{h\T}$ must not be too large. We expect that future
measurements can bring us exciting new insights on the spin of the
nucleon by including information of transverse momentum dependence
in data analysis.

\section*{Acknowledgement}
We are grateful to Harut Avakian, Wen Qian, Jacques Soffer, and Feng
Yuan for useful discussions. This work is partially supported by
National Natural Science Foundation of China (Nos.~10721063,
10575003, 10528510), by the Key Grant Project of Chinese Ministry of
Education (No.~305001), by the Research Fund for the Doctoral
Program of Higher Education (China).

\end{document}